\documentclass{article}

\usepackage{balance}

\usepackage{spconf,amsmath,graphicx}

\usepackage{bm, bbm}
\usepackage{graphicx}
\usepackage[margin = 2cm]{geometry}
\usepackage{multirow}

\newcommand{\T}{^\top}
\usepackage{empheq}
\usepackage{mathrsfs}
\usepackage{framed}

\usepackage{graphicx}
\usepackage{cite}
\usepackage{amsmath}
\usepackage{amssymb}
\usepackage{dsfont}
\usepackage[mathcal]{euscript}
\usepackage{empheq}
\usepackage{graphics}
\usepackage{bm}
\usepackage{psfrag}
\usepackage{mathrsfs}
\usepackage{bbm}
\usepackage{color}
\usepackage{cancel}
\input{epsf}
\newtheorem{theorem}{Theorem}
\newtheorem{lemma}{Lemma}

\newtheorem{assumption}{Assumption}

\newcommand{\beq}{\begin{equation}}
\newcommand{\eeq}{\end{equation}}
\newcommand{\beqa}{\begin{eqnarray}}
\newcommand{\eeqa}{\end{eqnarray}}
\newcommand{\dfz}{\triangleq}

\title{Learning Graph Influence from Social Interactions}

\name{Vincenzo Matta$^{\star}$ \qquad Virginia Bordignon$^{\dagger}$ \qquad Augusto Santos \qquad Ali H. Sayed$^{\dagger}$
\thanks{This work was supported in part by the Swiss National Science Foundation under grant 205121-184999. 
E-mails: vmatta@unisa.it, virginia.bordignon@epfl.ch, augusto.pt@gmail.com, ali.sayed@epfl.ch}
}

\address{$^{\star}$DIEM, University of Salerno. $^{\dagger}$EPFL, School of Engineering.}

\begin{document}
\ninept
\maketitle
\begin{abstract}
In social learning, agents form their opinions or beliefs about certain hypotheses by exchanging local information. 
This work considers the recent paradigm of {\em weak graphs}, where the network is partitioned into sending and receiving components, with the former having the possibility of exerting a domineering effect on the latter. 
Such graph structures are prevalent over social platforms. 
We will {\em not} be focusing on the direct social learning problem (which examines {\em what} agents learn), but rather on the dual or reverse learning problem (which examines {\em how} agents learned). Specifically, from observations of the stream of beliefs at certain agents, we would like to examine whether it is possible to learn the strength of the connections (influences) from sending components in the network to these receiving agents.
\end{abstract}

\section{Introduction and Related Work}
Social learning is a collective process where agents construct their individual beliefs about certain hypotheses by integrating the beliefs of neighboring agents into their own through social interaction~\cite{ChamleyBook,Jadbabaie2013,ScaglioneSPmag2013,ScaglioneACM2013}.
There exist several variants of social learning algorithms, which assume different protocols for the distributed propagation of information, as well as different ways of combining the neighbors' beliefs. 
Most algorithms rely either on consensus~\cite{Jad} or diffusion strategies~\cite{Zhao,YingSayed2016,Salami,NedicTAC2017,Javidi}, with some works using linear combination of beliefs~\cite{Jad,Zhao,YingSayed2016,Salami} and other works using logarithmic beliefs~\cite{NedicTAC2017,Javidi}. However, with the exception~\cite{Zhao,YingSayed2016,Salami}, most prior works focus mainly on {\em strongly-connected} networks (i.e., graphs where there is a direct and reverse path between any two agents, in addition to some agents having a self-loop as a sign of confidence in their own information).
Under this setting, the limiting (as time elapses) evolution of the individual agents' belief has been shown to converge collectively to the same opinion, which can be the true underlying hypothesis~\cite{Zhao,Jad,Javidi}, or a hypothesis minimizing a suitable objective function~\cite{NedicTAC2017}.   

The relevant case of {\em weakly-connected} networks has received less attention in the literature, despite its relevance for information spread over social platforms. Over weak graphs, some sub-components of the graph send information in one direction towards receiving agents but do not necessarily pay attention to (or even receive) information back. For example, a celebrity on Twitter may have thousands or millions of followers, but may be following only a handful of these individuals. Another example is media networks broadcasting information to a large number of users and hardly receiving feedback from any of these users.  More fundamentally, a weak graph is modeled as consisting of two components: {\em sending} sub-networks and {\em receiving} sub-networks~\cite{YingSayed2016, Salami,Zhao}. 
This paradigm was considered in~\cite{YingSayed2016, Salami} with reference to the linear-belief-combination rule, and in~\cite{MattaSantosSayedICASSP2019} for the log-belief combination rule. 
These works showed that, over weak graphs, receiving agents can be strongly influenced by one or more sending sub-networks independent of their own local observations. In this way, receiving agents can be made to move towards wrong beliefs by domineering sending sub-networks.

The earlier works focused on the direct learning problem. They examined the following fundamental question. 
Given a weak graph, how does information diffuse through the network and what are the limiting beliefs that agents converge to? Will they all converge to the same opinion or to scattered opinions? This paper examines the {\em dual} or reverse learning problem.  Assume we observe the evolution of the beliefs of certain receiving agents over time. Can we discover which sending sub-networks are most responsible for influencing the opinion formation of these agents? It is clear that this is a very relevant problem with many useful applications. It is also a challenging problem for reasons that will become clear as we progress with the presentation.

The inverse learning problem falls into the class of topology learning. However, in contrast with standard topology inference problems, there is an important element of novelty and distinction. In our formulation, we do {\em not} have access to the beliefs streaming from the sending agents. For this reason, we cannot exploit traditional methods where the estimation of connections between pairs of agents relies on comparison (e.g., correlation) between data streams coming from these {\em pairs} of agents~\cite{tomo,SantosMattaSayedIT2019,mateos}. 
We need to develop an alternative approach, which exploits to great effect our previous results on the direct learning problem. 
In particular, our analysis reveals a useful interplay between the two coexisting learning problems: $i)$ the {\em direct} inferential problem of social learning; and $ii)$ the {\em inverse} topology learning problem. 

\section{Background and Problem Setting}

A network of $N$ agents collects streaming data from the environment. 
The random variable $\bm{\xi}_{k,i}\in\mathcal{X}_k$  (we use bold fonts to emphasize randomness) describes the data at agent $k\in\{1,2,\ldots,N\}$ at time $i\in\mathbb{N}$. Data are independent over time, but they can be dependent across agents. 
The space $\mathcal{X}_k$ can vary across agents, and $\bm{\xi}_{k,i}$ is generated according to $f_k(\xi)$ (either a probability density function or a probability mass function), which is allowed to vary across agents as well. 
The goal of the agents is to choose one state of nature $\theta$ belonging to a finite set $\Theta=\{1,2,\ldots,H\}$. 
To accomplish this task, the agents assume a family of {\em likelihood functions} $L_k(\xi|\theta)$ with $\xi\in\mathcal{X}_k$. 
The dissimilarity between the true distribution $f_{k}(\xi)$ and the likelihood $L_k(\xi|\theta)$ is quantified through the Kullback-Leibler (KL) divergence $D[f_k || L_k(\theta)]$, which will be assumed finite for all agents and hypotheses.

Let us now describe the social learning strategy. 
Since at time $i=0$, the agents have no prior information to discard any hypothesis, we will assume that all agents assign nonzero probability mass to all hypotheses, namely, $\bm{\mu}_{k,0}(\theta)> 0$ for all $\theta\in\Theta$~\cite{NedicTAC2017,Javidi}. 
For any hypothesis $\theta\in\Theta$, agent $k$ at time $i$ employs its most recent {\em private} data, $\bm{\xi}_{k,i}$, to evaluate the likelihood $L_k(\bm{\xi}_{k,i}|\theta)$, which is in turn employed to update the local belief, $\bm{\mu}_{k,i-1}(\theta)$. This leads to an {\em intermediate} belief $\bm{\psi}_{k,i}(\theta)$ through the following Bayesian update:
\beq
\bm{\psi}_{k,i}(\theta)=
\displaystyle{
\frac
{\bm{\mu}_{k,i-1}(\theta)L_k(\bm{\xi}_{k,i} | \theta)}
{\displaystyle{\sum_{\theta^{\prime}\in\Theta} \bm{\mu}_{k,i-1}(\theta^{\prime})L_k(\bm{\xi}_{k,i} | \theta^{\prime})}}
}.
\label{eq:private}
\eeq
Second, in a combination step, agent $k$ aggregates the intermediate beliefs received from its neighbors by combining linearly the logarithm of these beliefs (exponentiation and normalization serve to give back an admissible belief):
\beq
\bm{\mu}_{k,i}(\theta)=
\displaystyle{
\frac{
\exp\left\{
\displaystyle{\sum_{\ell=1}^N a_{\ell k} \log \bm{\psi}_{\ell,i}(\theta)}
\right\}
}
{
\displaystyle{
\sum_{\theta^{\prime}\in\Theta} \exp\left\{
\sum_{\ell=1}^N a_{\ell k} \log \bm{\psi}_{\ell,i}(\theta^{\prime})
\right\}
}
}
}.
\label{eq:combine}
\eeq
The matrix $A=[a_{\ell k}]$ is left-stochastic since we assume that the weight $a_{\ell k}\geq 0$ is necessarily equal to zero if $k$ cannot receive data from $\ell$, and that the weights used by $k$ to scale the received beliefs from its neighbors add up to one.

We focus on the case of a weak graph, which is defined as follows~\cite{YingSayed2016,Salami}. 
The network $\mathcal{N}=\{1,2,\ldots N\}$ is divided into $S$ sending sub-networks (denoted by $\mathcal{N}_{s}$, for $s=1,2,\ldots,S$) and $R$ receiving sub-networks (denoted by $\mathcal{N}_{S+r}$, for $r=1,2,\ldots,R$):
\beq
\mathcal{N}=\mathcal{S}\cup\mathcal{R},\qquad
\mathcal{S}\dfz\bigcup_{s=1}^S \mathcal{N}_{s},\qquad
\mathcal{R}\dfz\bigcup_{r=1}^{R} \mathcal{N}_{S+r}.
\eeq
The combination matrix over weak graphs has the following block form (with increasing node ordering across the $S+R$ components):
\beq
A=\left[\begin{array}{c|c}
   A_{\mathcal{S}} & A_{\mathcal{S}\mathcal{R}}\\
   \hline
   0 & A_{\mathcal{R}}
\end{array}\right]
\label{eq:Ablockstruct}
\eeq
where the matrix $A_{\mathcal{S}}={\sf blockdiag}\left\{A_{\mathcal{N}_1},A_{\mathcal{N}_2},\ldots,A_{\mathcal{N}_S}\right\}$ contains the weights within the sending sub-networks, and has a block-diagonal form since communication between distinct sending sub-networks is not necessary (otherwise, sending sub-networks can be grouped into a larger sending sub-network).
Likewise, the left-bottom block of $A$ is zero since communication from receiving to sending sub-networks is forbidden. 
The $S$ sending sub-networks (resp., the $R$ receiving sub-networks) are all individually assumed {\em strongly connected} (resp., connected; meaning that self-loops are not necessary). Communication among the $R$ sub-networks is allowed. 
Finally, we assume that each receiving sub-network is connected to at least one sending agent.   

It was shown in~\cite{YingSayed2016} that the limiting combination matrix power has the following structure:
\beq
A_{\infty}\dfz\lim_{i\rightarrow\infty}A^i=
\left[
\begin{array}{c|c}
E & E W \\
\hline
0 & 0
\end{array}
\right]=
\left[
\begin{array}{c|c}
E & \Omega \\
\hline
0 & 0
\end{array}
\right],
\label{eq:limatweak}
\eeq
where $E={\sf blockdiag}\left\{p^{(1)}\mathbbm{1}^{\top}_{N_1},p^{(2)}\mathbbm{1}^{\top}_{N_2},\ldots,p^{(S)}\mathbbm{1}^{\top}_{N_S}\right\}$ is a block diagonal matrix that stacks the $N_s\times 1$ Perron eigenvectors $p^{(s)}$ associated with the $s$-th sending sub-network, $\mathbbm{1}_L$ is an all-ones vector of size $L\times 1$, and where:
\beq
W=A_{\mathcal{S}\mathcal{R}}\,(I - A_{\mathcal{R}})^{-1}, \quad
\Omega=E W.
\label{eq:omegadef}
\eeq
The entries of $\Omega$ are denoted by $[\omega_{\ell k}]$ and we keep indexing the columns of $\Omega$ with an index $k=|\mathcal{S}|+1,\ldots, |\mathcal{S}|+|\mathcal{R}|$.
Since the limiting matrix power is left-stochastic and has a zero right-bottom block, $\Omega$ is left-stochastic as well. 
From~(\ref{eq:omegadef}) we can also write $\Omega=E A_{\mathcal{S}\mathcal{R}}(I+A_{\mathcal{R}}+A_{\mathcal{R}}^2+\dots)$,
and we see that $\omega_{\ell k}$ embodies the sum of influences over all paths from sending agent $\ell$ to receiving agent $k$. 

Let us now introduce the following {\em average} divergence at {\em receiving} agent $k\in\mathcal{R}$:
\beq
\mathscr{D}_k(\theta)\dfz \sum_{\ell\in\mathcal{S}} \omega_{\ell k} D[f_{\ell}||L_{\ell}(\theta)],
\label{eq:avDiv}
\eeq
which is a weighted combination of the KL divergences pertaining {\em only} to the {\em sending} agents.
Throughout the work, we will invoke the following standard identifiability assumption.
\begin{assumption} (Unique Minimizer).
\label{assum:uniquemin}
For $k=1,2,\ldots,N$, the function $\mathscr{D}_k(\theta)$ has the unique minimizer: 
\beq
\theta^{\star}_k\dfz\arg\!\min_{\theta\in\Theta}\mathscr{D}_k(\theta).
\label{eq:uniquemin}
\eeq
~\hfill$\square$
\end{assumption}
It was shown in~\cite{MattaSantosSayedICASSP2019} that, under Assumption~\ref{assum:uniquemin}, the diffusion strategy in~(\ref{eq:private})--(\ref{eq:combine}) minimizes the divergence in~(\ref{eq:avDiv}), namely, that:
\beq
\lim_{i\rightarrow\infty}\bm{\mu}_{k,i}(\theta^{\star}_k)\stackrel{\textnormal{a.s.}}{=}1,
\label{eq:limbelief1st}
\eeq
where $\stackrel{\textnormal{a.s.}}{=}$ denotes almost-sure convergence.
Moreover, for all $\theta\neq \theta^{\star}_k$, the belief goes to zero exponentially as:
\beq
\lim_{i\rightarrow\infty}\frac{\log\bm{\mu}_{k,i}(\theta)}{i}
\stackrel{\textnormal{a.s.}}{=}
\mathscr{D}_k(\theta^{\star}_k)-\mathscr{D}_k(\theta).
\label{eq:divergencescondition}
\eeq

\section{Topology Learning}
In light of~(\ref{eq:uniquemin}), the particular opinion $\theta^{\star}_k$ that will be chosen by the $k$-th receiving agent is ultimately determined by the average divergence in~(\ref{eq:avDiv}). This dependence creates a strong tie between the network topology (through the limiting combination weights $\omega_{\ell k}$), and the shape of the beliefs.
We now examine the reverse problem. Assume the belief evolution of a receiving agent is monitored. 
This is a reasonable assumption since the information shared by the agents in the social learning strategy is actually constituted by the beliefs.
We want to use this information to infer the underlying links between the receiving agent and the sending sub-networks. 
This problem will be addressed under the following homogeneity assumption.
\begin{assumption}(Homogeneity in sending sub-networks).
\label{assum:homogeneity}
For $s=1,2,\ldots,S$, the distribution and the likelihood functions within the $s$-th sending sub-network are equal across all agents in that sub-network, namely, for all $\ell\in\mathcal{N}_s$:
\beq
f_{\ell}=f^{(s)},\qquad L_{\ell}(\theta)=L^{(s)}(\theta).
\eeq
~\hfill$\square$
\end{assumption}
Assumption~\ref{assum:homogeneity} implies that~(\ref{eq:avDiv}) becomes:
\beq
\mathscr{D}_k(\theta)=
\sum_{s=1}^S 
\left(D[f^{(s)} || L^{(s)}(\theta)] \sum_{\ell\in\mathcal{N}_s} \omega_{\ell k}\right),
\label{eq:homogeneity}
\eeq
which means that the topology influences $\mathscr{D}_k(\theta)$ only through an {\em aggregate} weight:
\beq
x_{sk}\dfz\sum_{\ell\in\mathcal{N}_s}\omega_{\ell k}=\sum_{\ell\in\mathcal{N}_s}w_{\ell k},
\label{eq:aggregateweightdef}
\eeq
where the latter equality comes from~(\ref{eq:omegadef}) and the definition of $E$. 
Now, while a weight $a_{\ell k}$ accounts for a {\em local} pairwise or {\em microscopic} interaction between $\ell$ and $k$, the aggregate weight $x_{sk}$ accounts for {\em macroscopic} topology effects, since: $i)$ $x_{sk}$ is determined by the {\em limiting} weights $\omega_{\ell k}$, which embody also effects {\em mediated} by multi-hop paths connecting $\ell$ and $k$; and $ii)$  $x_{sk}$ embodies the {\em global} effect coming from all agents belonging to the $s$-th sending component.  
Since we know that $\mathscr{D}_{k}(\theta)$ determines the behavior of the limiting belief, Eq.~(\ref{eq:homogeneity}) reveals that the topology ultimately determines the opinion chosen by a receiving agent only through the {\em global} weights $\{x_{sk}\}$. 

\begin{figure}[!t]
	\centering
	\includegraphics[width=.4\textwidth]{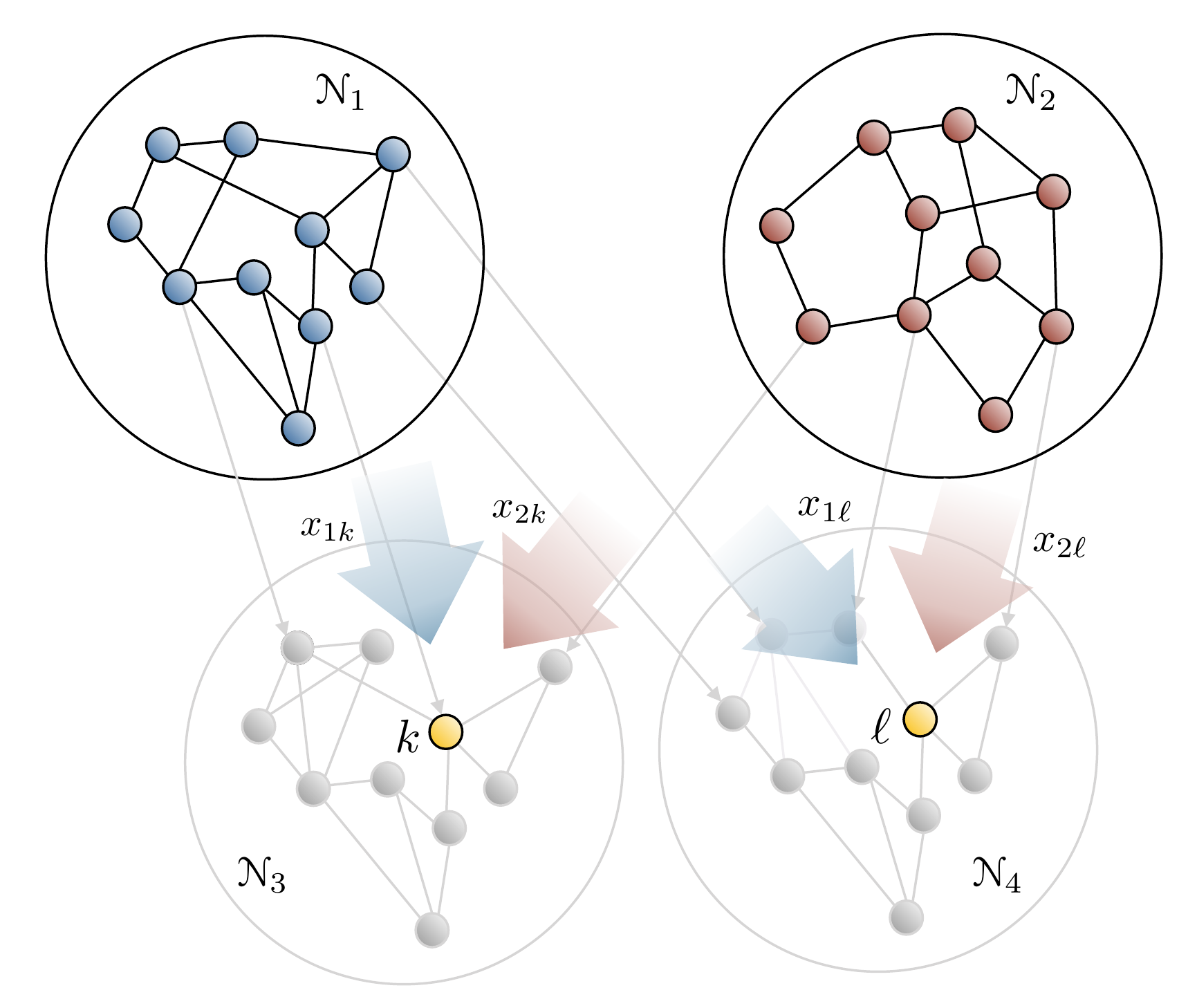}
	\caption{The topology inference goal is to estimate the {\em global} weights $x_{s k}$ linking sending sub-network $s$ to receiving agent $k$. The weight $x_{1k}$ in the figure embodies the influence of {\em all} sending agents in $\mathcal{N}_1$, from {\em all} paths (including intermediate receiving agents) leading to the receiving agent $k\in\mathcal{N}_3$.}
	\label{fig:macrotopology}
\end{figure}

Regarding the data used for topology inference, we assume the shared intermediate beliefs, $\bm{\psi}_{k,i}(\theta)$, are available. 
We will say that {\em consistent} topology learning is achievable if the $\{x_{sk}\}$ can be correctly estimated when sufficient time is given for learning. We focus accordingly on the {\em limiting} data:\footnote{We remark that $\bm{\psi}_{k,i}(\theta)$ and $\bm{\mu}_{k,i}(\theta)$ have the same limiting properties.}
\beq
y_k(\theta)\dfz\lim_{i\rightarrow\infty} \frac{\log\bm{\psi}_{k,i}(\theta)}{i}\stackrel{\textnormal{a.s.}}{=} 
\mathscr{D}_{k}(\theta^{\star}_k) - \mathscr{D}_{k}(\theta),
\label{eq:zklimdef}
\eeq
and formulate the following topology inference problem, which is illustrated in Fig.~\ref{fig:macrotopology}. 
Introduce the global-weight vector $x_k\dfz[x_{1k},x_{2k},\ldots,x_{Sk}]^{\top}$ and stack the $H$ limiting beliefs $y_k(\theta)$ as $y_k\dfz [y_k(1),y_k(2),\ldots,y_k(H)]^{\top}$. We would like to know whether we can estimate $x_k$ consistently from observation of $y_k$.

\begin{figure*}[!htb]
\begin{minipage}{.33\linewidth}
\vspace*{-20pt}
{\centering{\bf ~~~~~~network graph}\par\medskip}
\centering
{
\includegraphics[scale=.43]{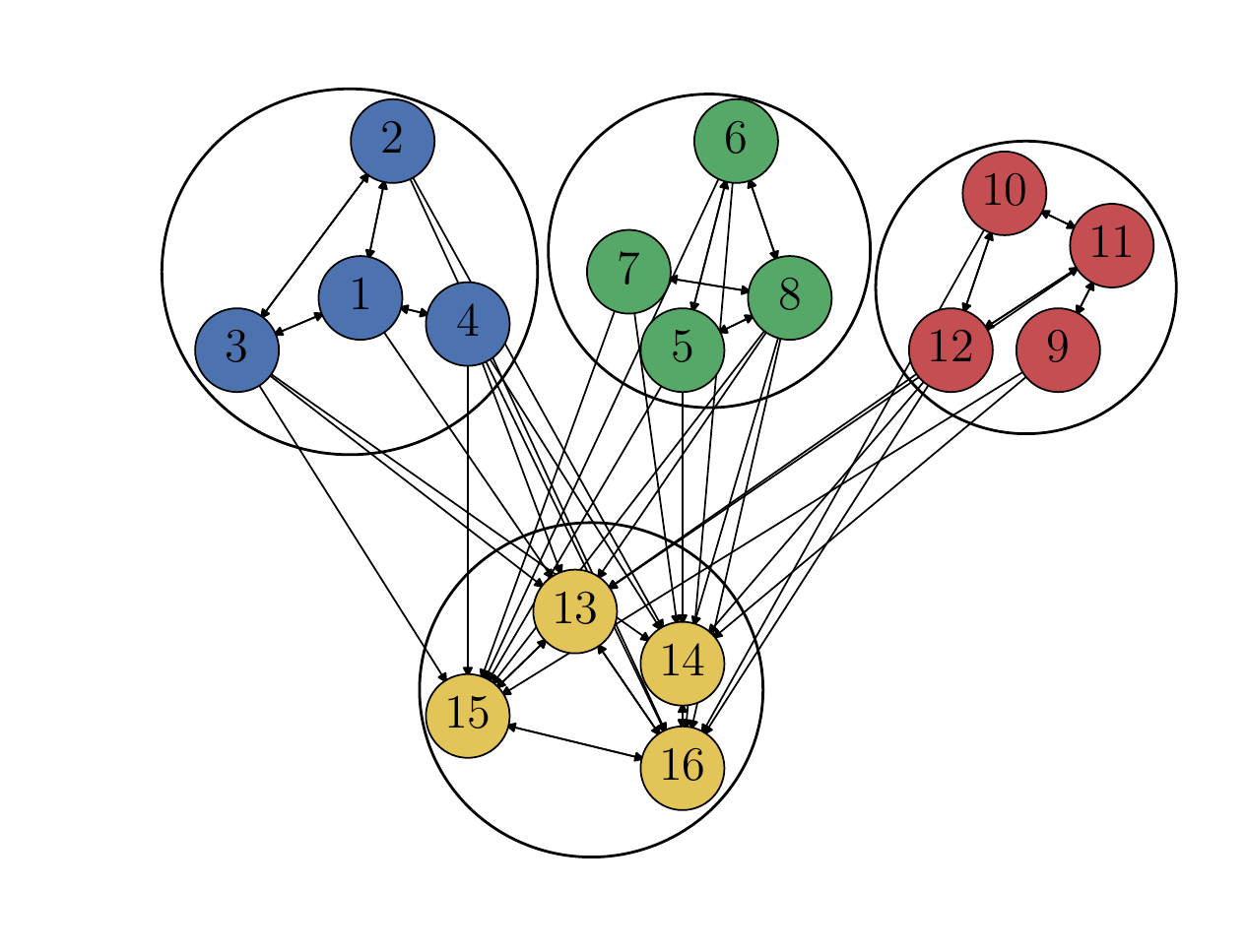}}
\end{minipage}
\begin{minipage}{.33\linewidth}
{\centering{\bf ~~~~~~belief evolution}\par\medskip}
\centering
{
\includegraphics[scale=.43]{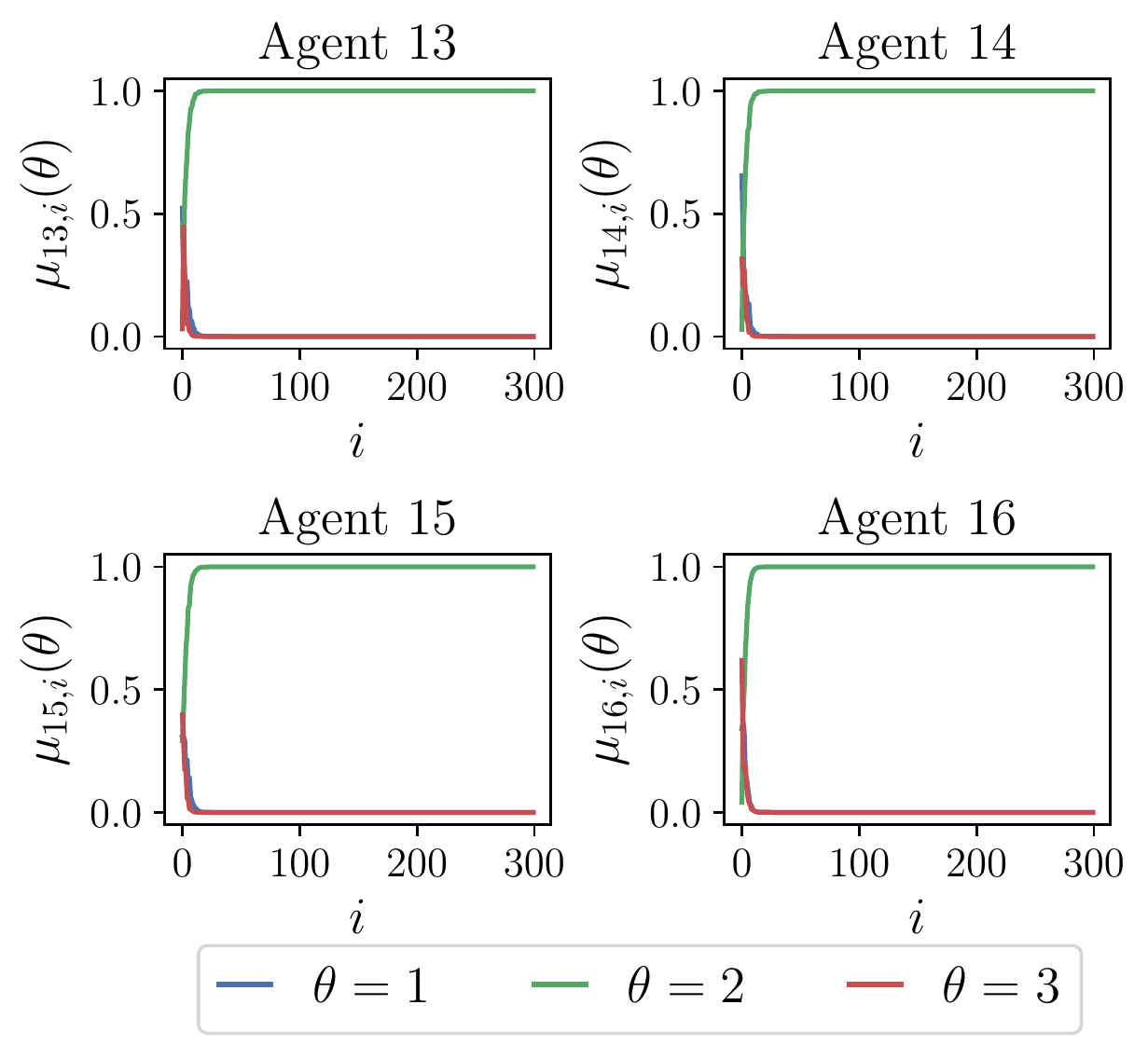}}
\end{minipage}
\begin{minipage}{.33\linewidth}
{\centering{\bf ~~~~~~estimated weights}\par\medskip}
\centering
{
\includegraphics[scale=.43]{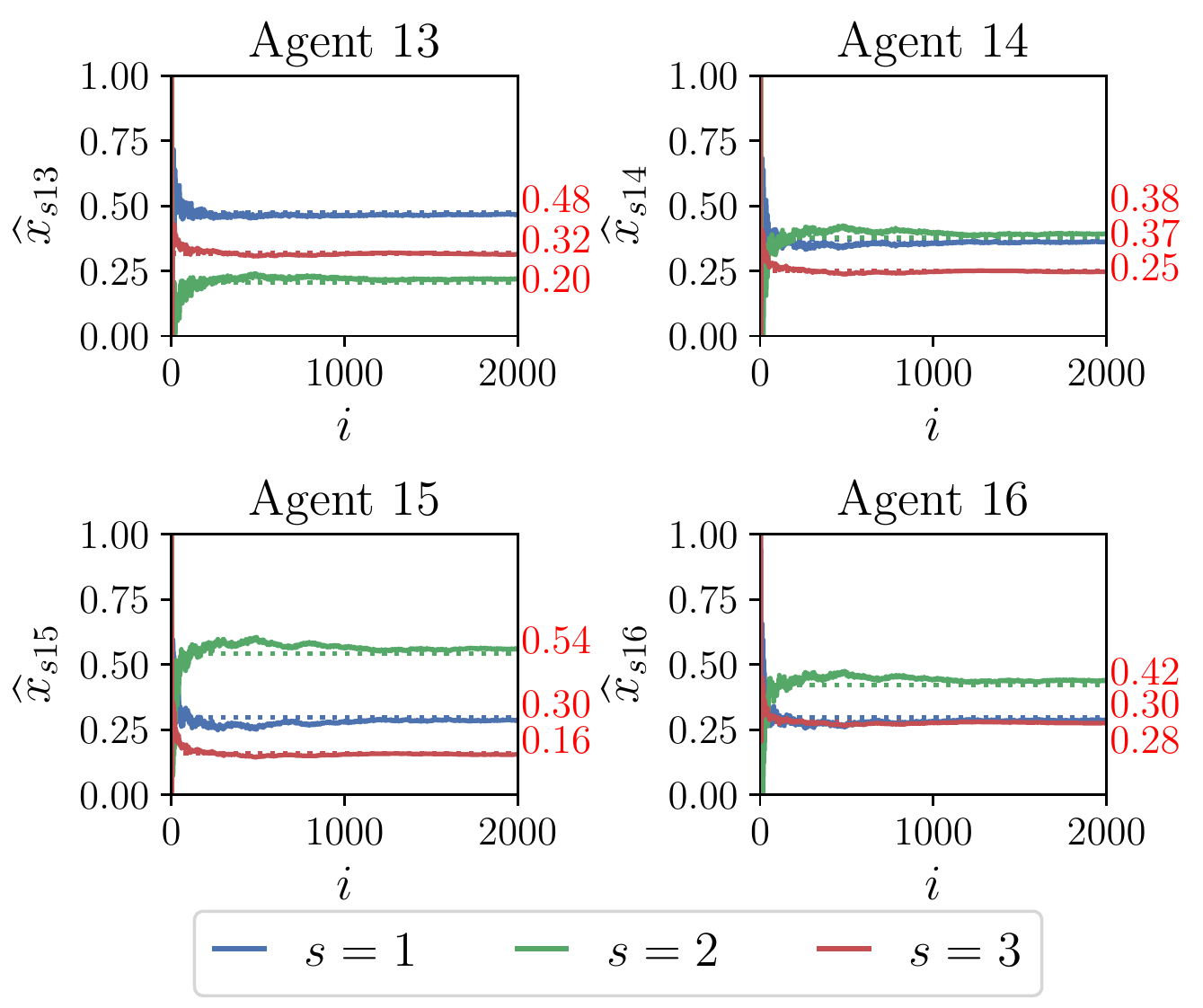}}
\end{minipage}
\caption{Randomly perturbed Gaussian model. {\em Leftmost panel}. Network topology. {\em Middle panel}. Belief convergence of receiving agents. {\em Rightmost panel}. Estimated limiting topology, with the red numbers denoting the true values $\{x_{sk}\}$.}
\label{fig:simTheorem3Gauss}
\end{figure*}

It is useful to introduce the $H\times S$ matrix $[D]_{\theta s}=d_{\theta s}=D[f^{(s)}||L^{(s)}(\theta)]$, which allows rewriting the limiting data as:
\beq
y_k(\theta)=\mathscr{D}(\theta^{\star}_k)-\mathscr{D}(\theta)=\sum_{s=1}^S (d_{\theta^{\star}_k s} - d_{\theta s}) \, x_{sk}.
\eeq
Accordingly, we see that the topology inference problem can be recast in terms of the following constrained linear system: 
\beq
\textnormal{Find }\widetilde{x}_k\in\mathbb{R}^S: ~~\widetilde{y}_k=C_k\, \widetilde{x}_k,~~~\widetilde{x}_k>0,
\label{eq:linsystopology}
\eeq 
where we defined:
\beq
B_k\dfz \left(\mathbbm{1}_{H} e\T_{\theta^{\star}_k} - I_H\right) D,\quad
C_k
\dfz 
\begin{bmatrix}
B_k
\\
\mathbbm{1}_S\T
\end{bmatrix},\quad
\widetilde{y}_k\dfz
\begin{bmatrix}
y_k
\\
1
\end{bmatrix},
\label{eq:augmentedB}
\eeq
with $e_m$ being an $H\times 1$ vector with all zeros and a one in the $m$-th position, and where the last row in $C_k$ and the last entry in $\widetilde{y}_k$ serve to embody the convexity constraint $\sum_{s=1}^S \widetilde{x}_{sk}=1$. 
We want to examine the achievability of consistent topology learning. We study this problem under the assumption that the matrices $D$ and $B_k$ are known.\footnote{$B_k$ depends on $\theta^{\star}_k$, which can be estimated consistently from $y_k(\theta)$.  
} 
Now, achievability of consistent topology learning translates into the condition that the linear system in~(\ref{eq:linsystopology}) admits a unique solution. We observe that the augmented matrix $C_k$ is an $(H+1)\times S$ matrix with an all-zeros row. Under the assumption that the global weight vector does not contain zeros, the following lemma can be proved (proof omitted due to space limitations).
\begin{lemma}({\bf Necessary Condition for Topology Learning}).
\label{prop:necessary}
The system in~(\ref{eq:linsystopology}) admits a unique solution if, and only if, $\mathrm{rank}(C_k)=S$.
Thus, a necessary condition for topology learning is:
\beq
H\geq S,
\eeq  
i.e., that the number of hypotheses is at least equal to the number of sending sub-networks.~\hfill$\blacksquare$
\end{lemma}

Lemma~\ref{prop:necessary} reveals a remarkable interplay between Social Learning (SL) and Topology Learning (TL).
One interpretation of the condition $H \geq S$ is that the TL problem becomes feasible when its complexity (number of sending components) is not greater than the SL complexity (number of hypotheses). 
Lemma~\ref{prop:necessary} reveals also that TL consistency is not easily granted. For example, if the agents want to solve a binary detection problem, the maximum number of sending sub-networks that could allow consistent TL is $S=2$. 

\subsection{Structured Gaussian Models}
\label{sec:TLGauss}
We now introduce a useful Gaussian model that can arise in many applications. 
We assume that all agents use the same family of likelihoods $\{L(\theta)\}$, for $\theta=1,2,\ldots,H$, which are unit-variance Gaussian likelihoods with different means $\{\mathsf{m}_\theta\}$. 
Each true distribution coincides with one of the likelihoods, which means that $f^{(s)}$ is a unit-variance Gaussian distribution with mean $\nu_s$ that is chosen among the means $\{\mathsf{m}_\theta\}$, namely, $\nu_s\in\{\mathsf{m}_1,\mathsf{m}_2,\ldots,\mathsf{m}_H\}$. The sending sub-networks have different means. 
Without loss of generality, we assume that the sending sub-networks are numbered so that the means of the true distributions are $\nu_1=\mathsf{m}_1, \ldots,\nu_S=\mathsf{m}_S$, which implies that the divergence matrix $D$ is equal to:
\beq
\frac 1 2
\begin{bmatrix}
	0&(\mathsf{m}_1-\mathsf{m}_2)^2&\dots &(\mathsf{m}_1-\mathsf{m}_S)^2\\
	(\mathsf{m}_2-\mathsf{m}_1)^2&0&\dots &(\mathsf{m}_2-\mathsf{m}_S)^2\\
	\vdots&&&\vdots\\
	(\mathsf{m}_H-\mathsf{m}_1)^2&(\mathsf{m}_H-\mathsf{m}_2)^2&\dots &(\mathsf{m}_H-\mathsf{m}_S)^2
	\end{bmatrix}.
	\label{eq:EDMGauss}
\eeq
For $H=S$, the matrix $D$ is a Euclidean distance matrix (but for the constant $1/2$)~\cite{dokmanic2015euclidean}. 
These matrices are constructed as follows. 
Given points $r_1,r_2,\ldots,r_L$, belonging to $\mathbb{R}^{\sf dim}$, the $(i,j)$-th entry of the matrix ${\sf EDM}(r_1,r_2,\ldots,r_L)$ is given by the squared Euclidean distance between points $r_i$ and $r_j$. 
We see then from~(\ref{eq:EDMGauss}) that, for $H=S$:
\beq
D=\frac 1 2 {\sf EDM}(\mathsf{m}_1,\mathsf{m}_2,\ldots,\mathsf{m}_H).
\eeq
In the case $H>S$, the matrix $D$ can be described as an {\em extended} Euclidean distance matrix: 
\beq
D=\begin{bmatrix}
E_S\\
F
\end{bmatrix},\qquad
E_{H}=\begin{bmatrix}
E_S&F\T\\
F&E_{H-S}
\end{bmatrix},
\label{eq:generalD}
\eeq
where:
\beqa
E_S&\dfz& \frac 1 2 {\sf EDM}(\mathsf{m}_1,\mathsf{m}_2,\ldots,\mathsf{m}_S),\nonumber\\
E_H&\dfz& \frac 1 2 {\sf EDM}(\mathsf{m}_1,\mathsf{m}_2,\ldots,\mathsf{m}_H),\nonumber\\
E_{H-S}&\dfz& \frac 1 2 {\sf EDM}(\mathsf{m}_{S+1},\mathsf{m}_{S+2},\ldots,\mathsf{m}_H),
\label{eq:variousEDMdef}
\eeqa
and where $F$ is the $(H-S)\times S$ matrix with entries, for $\theta=S+1,S+2,\ldots,H$ and $s=1,2,\ldots,S$:
\beq
[F]_{\theta s}=\frac 1 2 (\mathsf{m}_\theta-\mathsf{m}_s)^2.
\eeq
The following theorem ascertains the feasibility of the TL problem for the structured Gaussian model. The proof relies heavily on some fundamental properties of Euclidean distance matrices, and is omitted for space limitations. 
\begin{theorem} ({\bf Topology Learning under Structured Gaussian Models}).
\label{theor:TopologyGaussian}
Let $S\geq 2$ and $H\geq S$. 
Under the structured Gaussian model and Assumption~\ref{assum:uniquemin}, for all $k\in\mathcal{R}$ we have that $\mathrm{rank}(C_k)=2$~\hfill$\blacksquare$
\end{theorem}

In view of Lemma~\ref{prop:necessary}, Theorem~\ref{theor:TopologyGaussian} implies that under the structured Gaussian model topology learning is very challenging, as it is feasible only when $S=2$. 

\subsection{Diversity Models}
\label{sec:TLrandomized}
Once ascertained that the topology over a structured Gaussian model is difficult to learn, we now examine the effect that diversity in the models of the sending sub-networks can have on TL.
Differently from the previous section, we require that the entries of $D$ are not tightly related, and we allow them to assume values in $\mathbb{R}_{+}^{H\times S}$ ($\mathbbm{R}_+$ collects the nonnegative reals) with no structure linking them.   
As a formal way to embody this degree of variability in how the agents ``see'' the world, we model the divergences as jointly absolutely continuous random variables (bold notation $\bm{d}_{\theta s}$). 
Under this framework, it is possible to establish the following result, whose proof is omitted for space constraints.

\begin{theorem}({\bf Topology Learning under General Models with Diversity}).
\label{theor:TopologyRandom}
Assume that the array $\{\bm{d}_{\theta s}\}$ is made of random variables that are jointly absolutely continuous w.r.t. the Lebesgue measure on $\mathbbm{R}_{+}^{H\times S}$. 
If $H\geq S$, Assumption~\ref{assum:uniquemin} is verified and the matrix $\bm{C}_k$ is full column rank with probability $1$, for all $k\in\mathcal{R}$.~\hfill$\blacksquare$
\end{theorem}

Theorem~\ref{theor:TopologyRandom} reveals that {\em divergence configurations leading to a rank-deficient matrix $C_k$ are rare} if sufficient diversity exists in the models of the sending components, i.e., the TL problem is feasible for most configurations.

\section{Illustrative Example}
We show an example pertaining to Theorem~\ref{theor:TopologyRandom}, for a case with $H=S=3$. 
The network topology is shown in the leftmost panel of Fig.~\ref{fig:simTheorem3Gauss}. 
The true distribution of sub-network $s=1,2,3$ is a unit-variance Gaussian with mean $s$. 
The likelihood of the $s$-th sending sub-network, evaluated at hypothesis $\theta$, is unit-variance Gaussian with expectation $\theta+\bm{u}_{\theta s}$, with $\bm{u}_{\theta s}$ being independent random variables uniformly distributed in $[-0.1, 0.1]$.
The middle panel of Fig.~\ref{fig:simTheorem3Gauss} pertains to the SL problem, as it displays the convergence of the receiving agents' beliefs. 
In the considered example, sub-network $s=2$ (green agents) exerts a domineering role, since the beliefs of the receiving agents converge to opinion $\theta=2$.

We move on to the TL problem. First, for an observation time $i$, we construct the empirical data $\widehat{y}_k(\theta)=(1/i)\log \bm{\psi}_{k,i}(\theta)$, and estimate $\theta^{\star}_k$ as the value that maximizes $\widehat{y}_k(\theta)$.
Then, we solve~(\ref{eq:linsystopology}) with empirical matrices replacing the exact ones to estimate the connection-weight vector $x_{k}$. Provided that the system evolves for a sufficiently long time, this procedure allows to retrieve the true $x_k$, as shown in the rightmost panel of Fig.~\ref{fig:simTheorem3Gauss}.

\vspace*{-10pt}
\section{Conclusion}
\label{sec:SLvsTL}
This work considered the following {\em dual} problem of social learning over weakly-connected networks. 
Given observation of {\em what} the agents are learning (Social Learning, SL), we want to discover {\em how} they are being influenced from the sending agents (Topology Learning, TL). 
We established that a necessary condition for consistent TL is that the number of hypotheses $H$ is at least equal to the number of sending components $S$. In other words, the complexity of the TL problem (number of sub-networks) must be not greater than the complexity of the SL problem (number of hypotheses).   
We examined two models. A structured Gaussian model where {\em all} sending sub-networks use the same family of Gaussian likelihoods, and the true distributions are chosen within this family and are distinct across the sending sub-networks. 
We showed that for this model TL is feasible only when $S=2$, due to the limited diversity across the sending sub-networks.
Accordingly, we examined another model, where the likelihoods and the true distributions exhibit a certain {\em diversity}. 
For this case, we showed that the TL problem is feasible with probability one provided that $H\geq S$.
In summary, the two critical features to enable consistent TL are: {\em more hypotheses than sending components} and {\em a sufficient degree of diversity}.

\balance
\vfill
\pagebreak


\begin{thebibliography}{10}
\providecommand{\url}[1]{#1}
\csname url@samestyle\endcsname
\providecommand{\newblock}{\relax}
\providecommand{\bibinfo}[2]{#2}
\providecommand{\BIBentrySTDinterwordspacing}{\spaceskip=0pt\relax}
\providecommand{\BIBentryALTinterwordstretchfactor}{4}
\providecommand{\BIBentryALTinterwordspacing}{\spaceskip=\fontdimen2\font plus
\BIBentryALTinterwordstretchfactor\fontdimen3\font minus
  \fontdimen4\font\relax}
\providecommand{\BIBforeignlanguage}[2]{{%
\expandafter\ifx\csname l@#1\endcsname\relax
\typeout{** WARNING: IEEEtran.bst: No hyphenation pattern has been}%
\typeout{** loaded for the language `#1'. Using the pattern for}%
\typeout{** the default language instead.}%
\else
\language=\csname l@#1\endcsname
\fi
#2}}
\providecommand{\BIBdecl}{\relax}
\BIBdecl

\bibitem{ChamleyBook}
C.~Chamley, \emph{Rational Herds: Economic Models of Social Learning}.\hskip
  1em plus 0.5em minus 0.4em\relax Cambridge, UK: Cambridge Univ. Press, 2004.

\bibitem{Jadbabaie2013}
A.~Jadbabaie, P.~Molavi, and A.~Tahbaz-Salehi, ``Information heterogeneity and
  the speed of learning in social networks,'' \emph{Columbia Business School
  Research Paper}, pp. 13--28, May 2013.

\bibitem{ScaglioneSPmag2013}
C.~Chamley, A.~Scaglione, and L.~Li, ``Models for the diffusion of beliefs in
  social networks: An overview,'' \emph{{IEEE} Signal Process. Mag.}, vol.~30,
  no.~3, pp. 16--29, May 2013.

\bibitem{ScaglioneACM2013}
E.~Yildiz, A.~Ozdaglar, D.~Acemoglu, A.~Saberi, and A.~Scaglione, ``Binary
  opinion dynamics with stubborn agents,'' \emph{ACM Trans. Econ. Comput.},
  vol.~1, no.~4, pp. 19:1--19:30, Dec. 2013.

\bibitem{Jad}
A.~Jadbabaie, P.~Molavi, A.~Sandroni, and A.~Tahbaz-Salehi, ``Non-{B}ayesian
  social learning,'' \emph{Games and Economic Behavior}, vol.~76, no.~1, pp.
  210--225, Sep. 2012. 
  
\bibitem{Zhao}
X.~Zhao and A.~H. Sayed, ``Learning over social networks via diffusion
  adaptation,'' in \emph{Proc. Asilomar Conference on Signals, Systems and
  Computers}, Nov. 2012, pp. 709--713.

\bibitem{YingSayed2016}
B.~Ying and A.~H. Sayed, ``Information exchange and learning dynamics over
  weakly connected adaptive networks,'' \emph{{IEEE} Trans. Inf. Theory},
  vol.~62, no.~3, pp. 1396--1414, Mar. 2016.

\bibitem{Salami}
H.~Salami, B.~Ying, and A.~H. Sayed, ``Social learning over weakly connected
  graphs,'' \emph{{IEEE} Trans. Signal Inf. Process. Netw.}, vol.~3, no.~2, pp.
  222--238, Jun. 2017.

\bibitem{NedicTAC2017}
A.~Nedi\'{c}, A.~Olshevsky, and C.~A. Uribe, ``Fast convergence rates for
  distributed non-{B}ayesian learning,'' \emph{{IEEE} Trans. Autom. Control},
  vol.~62, no.~11, pp. 5538--5553, Nov. 2017.

\bibitem{Javidi}
A.~Lalitha, T.~Javidi, and A.~D. Sarwate, ``Social learning and distributed
  hypothesis testing,'' \emph{{IEEE} Trans. Inf. Theory}, vol.~64, pp.
  6161--6179, Sep. 2018.

\bibitem{MattaSantosSayedICASSP2019}
V.~Matta, A.~Santos, and A.~H. Sayed, ``Exponential collapse of social beliefs over weakly-connected heterogeneous networks,'' in \emph{Proc. IEEE ICASSP}, Brighton, UK, May 2019, pp. 5267--5271.

\bibitem{tomo}
V.~Matta and A.~H. Sayed, ``Consistent tomography under partial observations over adaptive networks,'' \emph{IEEE Trans. Inf. Theory}, vol.~65, no.~1, pp. 622--646, Jan. 2019.

\bibitem{SantosMattaSayedIT2019}
A.~Santos, V.~Matta, and A.~H. Sayed, ``Local tomography of large networks under the low-observability regime,'' \emph{IEEE Trans. Inf. Theory}, available in early access, Oct. 2019, doi: 10.1109/TIT.2019.2945033.

\bibitem{mateos}
G.~Mateos, S.~Segarra, A.~Marques, and A.~Ribeiro, ``Connecting the dots: Identifying network structure via graph signal processing,'' \emph{IEEE Signal Process. Mag.}, vol.~36, no.~3, pp. 16--43, May 2019.

\bibitem{dokmanic2015euclidean}
I.~Dokmanic, R.~Parhizkar, J.~Ranieri, and M.~Vetterli, ``Euclidean distance
  matrices: Essential theory, algorithms, and applications,'' \emph{IEEE Signal Process. Mag.}, vol.~32, no.~6, pp. 12--30, Nov. 2015.

\end{thebibliography}


\end{document}